\documentclass[a4paper,11pt]{article}
\usepackage[left=2.5cm,top=3cm,right=2.5cm,bottom=3cm,bindingoffset=0.0cm]{geometry}

\usepackage[utf8]{inputenc}  
\usepackage[T1]{fontenc}     
\usepackage{amsmath, amssymb}
\usepackage{graphicx}        
\usepackage{hyperref}        
\usepackage[backend=biber,
            style=numeric,
            datamodel=standard,
            sorting=none,
            doi=true,
            url=false,
            isbn=false]{biblatex}
\title{Residence-time theory applied to circulating-fuel reactors: zero-power analysis}

\author{\textbf{Lubomír Bureš}\\[4pt] Saltfoss Energy ApS, Titangade 11, 2200 Copenhagen, Denmark \\[4pt] \texttt{lubomir.bures@saltfoss.com}}

\date{July 9, 2025}

\begin{document}
\maketitle

\begin{abstract}
  \noindent Circulating-fuel reactors (CFRs) lose reactivity when delayed-neutron precursors (DNPs) drift out of the core and may regain part of it when the fuel re-enters the core. This paper formulates a physics-based description of both effects by combining DNP transport with residence-time theory. Then, treating the core and ex-core regions as two mixing volumes in series, closed-form expressions for (i) the static reactivity loss due to precursor drift and (ii) the zero-power transfer function that governs linearised dynamics are derived. When the gamma residence-time distributions are used, the new framework is shown to reduce to the plug-flow and Continuous-Stirred-Tank-Reactor limits as special cases, while generalising to intermediate mixing regimes via a single parameter: the degree of mixing. Performed parameter studies show that DNP recirculation has the highest impact when core and ex-core residence times are comparable and the product of the DNP decay constant and the in-core residence time is small. Benchmarks against the Molten-Salt Reactor Experiment are able to reproduce the measured static loss ($k_0 \approx 0.32$ \$) and its frequency response, with $\approx$20\% of the steady-state DNP worth arising from recirculation. Additionally, for the EVOL reference Molten-Salt Fast Reactor the model is shown to agree well with the results of high-fidelity Serpent-2 calculations coupled with Computational Fluid Dynamics. Overall, the residence-time approach offers a computationally light yet versatile tool for sensitivity studies and generation of physical intuition for the behaviour of CFRs. Foundation for extensions to importance weighting of DNPs and application of the framework to time-domain analysis is also briefly sketched.

\end{abstract}

\section{Introduction}
\label{sec:intro}
In circulating-fuel reactors (CFRs), delayed neutron precursors (DNPs) are advected with the flow of the fuel. As a result, the emission of a delayed neutron resulting from a DNP decay occurs at a different location than the fission event in which the precursor was born. Some of the DNPs decay in areas of low neutron importance outside of the active core, for example in the piping of the primary circuit. From the perspective of the chain reaction, the emitted delayed neutrons are effectively lost and this leads to the reduction of the nuclear reactivity of the CFR. This phenomenon is known as the ``static reactivity loss due to precursor drift'' \cite{Akcasu1971,Aufiero2014} and is of high relevance for the safety and stability of the system. In simplified terms, the evolution of a reactor transient is dependent on reactivity in dollars $k$, i.e.\ $\rho$ (-) scaled by the delayed neutron fraction $\beta$ (-) \cite{Akcasu1971}. As all neutrons lost due to precursor drift are by definition delayed, one could crudely consider the delayed neutron fraction of a CFR to be equal to $(\beta-\rho_0)$, where $\beta$ would be the delayed neutron fraction under static-fuel conditions and $\rho_0$ the reactivity loss due to precursor drift under circulating conditions. This means that CFRs are in general more responsive to reactivity perturbations, as can be seen e.g.\ from the reactivity-to-power transfer functions \cite{Cammi2010,Paszit2017,Bures2024a,Bures2024b}.

The evaluation of the static reactivity loss is made even more complex due to the fact that some of the DNPs that are advected from the active core return due to recirculation. That has further implications for the analysis of the dynamic behaviour of CFRs by introducing non-trivial physical coupling between the various zones of the reactor system. In general, when performing a theoretical or computational analysis of a CFR, it is necessary to consider the motion of DNPs through both the reactor and the rest of the primary circuit. It has been shown in past work that the assumptions on flow behaviour strongly affect the results of the analysis and the treatment of DNP transport as simple 1D advection problem of the plug-flow type:
\begin{equation}
	\frac{\partial C}{\partial t} + U\frac{\partial C}{\partial x} = S,
	\label{eq:advection}
\end{equation}
can often lead to incorrect results. For example, in \cite{PonceTovar2024}, the effect of velocity profile was studied and it was found that assuming either plug flow or parabolic flow in the core can lead to changes in the estimated precursor drift. In \cite{Bures2024b}, a detailed comparison was made between the results of a pure plug-flow model and a model considering perfect mixing in the reactor core, showing clear differences in precursor drift and reactor frequency response. In some works, modifications to the advection model were proposed to address the issues associated with pure plug-flow modelling; for example in \cite{Kerlin1971} a mixing pot was added at the outlet in a dynamic model of the Molten Salt Reactor Experiment (MSRE). In \cite{Morgan2018}, the advection equation \eqref{eq:advection} was augmented by the inclusion of turbulent dispersion effects.

It should be noted that the historical MSRE dynamic model incorporated inputs from experimental measurements with the earlier predictions being in disagreement \cite{Kerlin1971,Haubenreich1962}. This indicates the need for simplified models to be informed by reference data. They can be obtained either from an experiment or, in modern times, potentially also from simulations performed with high-fidelity, multi-physics tools, such as GeN-Foam \cite{Shi2021} or the MOOSE framework \cite{AbouJaoude2021}. Nevertheless, even if high-fidelity models are available, development of simplified, physics-informed methods is still valuable, as they help to generate insights into the behaviour of the system and can be readily used for parameter sweeps and sensitivity studies.

Furthermore, the phenomenology of CFRs, in particular molten-salt reactors, is rather involved and many
physico-chemical phenomena become relevant when a system-wide analysis is to be performed. Considering the finite-resource constraints accompanying any real-world development and application of high-fidelity simulation tools, identified phenomena must be compared in importance before they are introduced into a simulation suite. If it is found that the impact of a phenomenon on the behaviour of the system is minor, it can be potentially neglected, reducing the overall complexity of the model. This is the foundation for developing robust, representative models that can be used to improve the understanding of the system and its behaviour under normal and off-normal conditions. This can also be achieved through the deployment and use of simplified models.

Thus, development and application of a simplified theoretical model for CFR dynamics, in particular zero-power transfer functions and static reactivity loss due to precursor drift, is the topic of this paper. To this end, the residence time theory is used. The rest of the paper is organised as follows: in Section \ref{sec:theor}, the theoretical model is derived. Then, in Section \ref{sec:res}, quantitative analysis of static reactivity loss due to DNP drift and zero-power transfer function using the developed model is performed with added focus on the role of recirculation. Comparisons are made against reference experimental and high-fidelity numerical data. Finally, conclusions are drawn in Section~\ref{sec:conc}.

N.B.\ Preliminary results of this work were presented at the 2025 International Conference on Mathematics and Computational Methods Applied to Nuclear Science and Engineering (M\&C 2025) \cite{Bures2025}. 


\section{Theoretical derivation}
\label{sec:theor}

\begin{figure}
	\centering
	\includegraphics[width=0.65\linewidth]{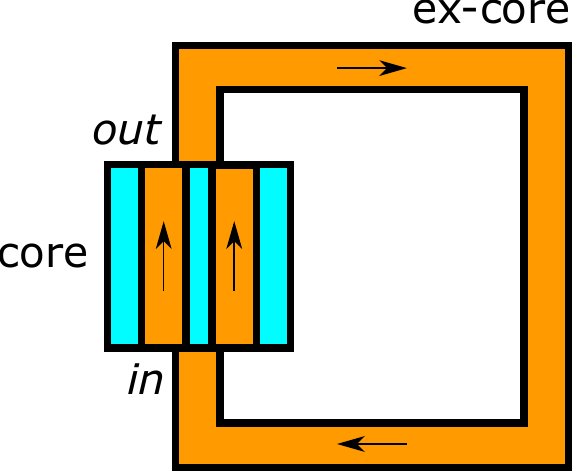}
	\caption{Schematic layout of the system considered in this work. The core region $c$ is indicated with the rest of the system being labelled as ex-core $e$. Inlet and outlet locations of the core are indicated and arrows are included to show the direction of the fuel flow.}
	\label{fig:cstr}
\end{figure}

\begin{figure}
	\centering
	\includegraphics[width=0.8\linewidth]{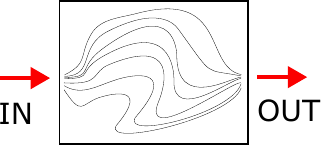}
	\caption{Schematic representation of a control volume $\mathcal{V}$ with inlet and outlet indicated.}
	\label{fig:cv}
\end{figure}

In this section, the zero-power transfer function and static reactivity loss due to precursor drift are derived for a simplified representation of a circulating-fuel reactor system shown in Fig.\ \ref{fig:cstr} using residence-time theory. In this system, it is considered that fission (and therefore production of DNPs) takes place only in the core region.

\subsection{Material balance equations}
A CFR system can be thought to be comprised by a network of connected control volumes; the embodiment shown in Fig.\ \ref{fig:cstr} consists of two control volumes, the core region $c$ and the ex-core region $e$, which are connected together to form a loop, i.e.\ the inlet of one of the regions is the outlet of the other and vice versa. Control volumes $\mathcal{V}$ with a single inlet and outlet can be represented schematically as shown in Fig.\ \ref{fig:cv}. The fluid fuel is assumed to be flowing through $\mathcal{V}$ from the inlet to the outlet and carrying tracers, which have a certain concentration $C_{\textit{in}}$ at the inlet and $C_{\textit{out}}$ at the outlet; these concentrations are in general functions of time $t$ (s). It is also assumed here that the fluid mass density is constant and the flow is steady, i.e.\ volumetric flow rate $Q$ (m$^3$/s) is constant as well. The units of the tracer concentration are taken for the moment to be arbitrary per unit volume, i.e.\ a.u./m$^3$.

\begin{figure}
	\centering
	\includegraphics[width=0.6\linewidth]{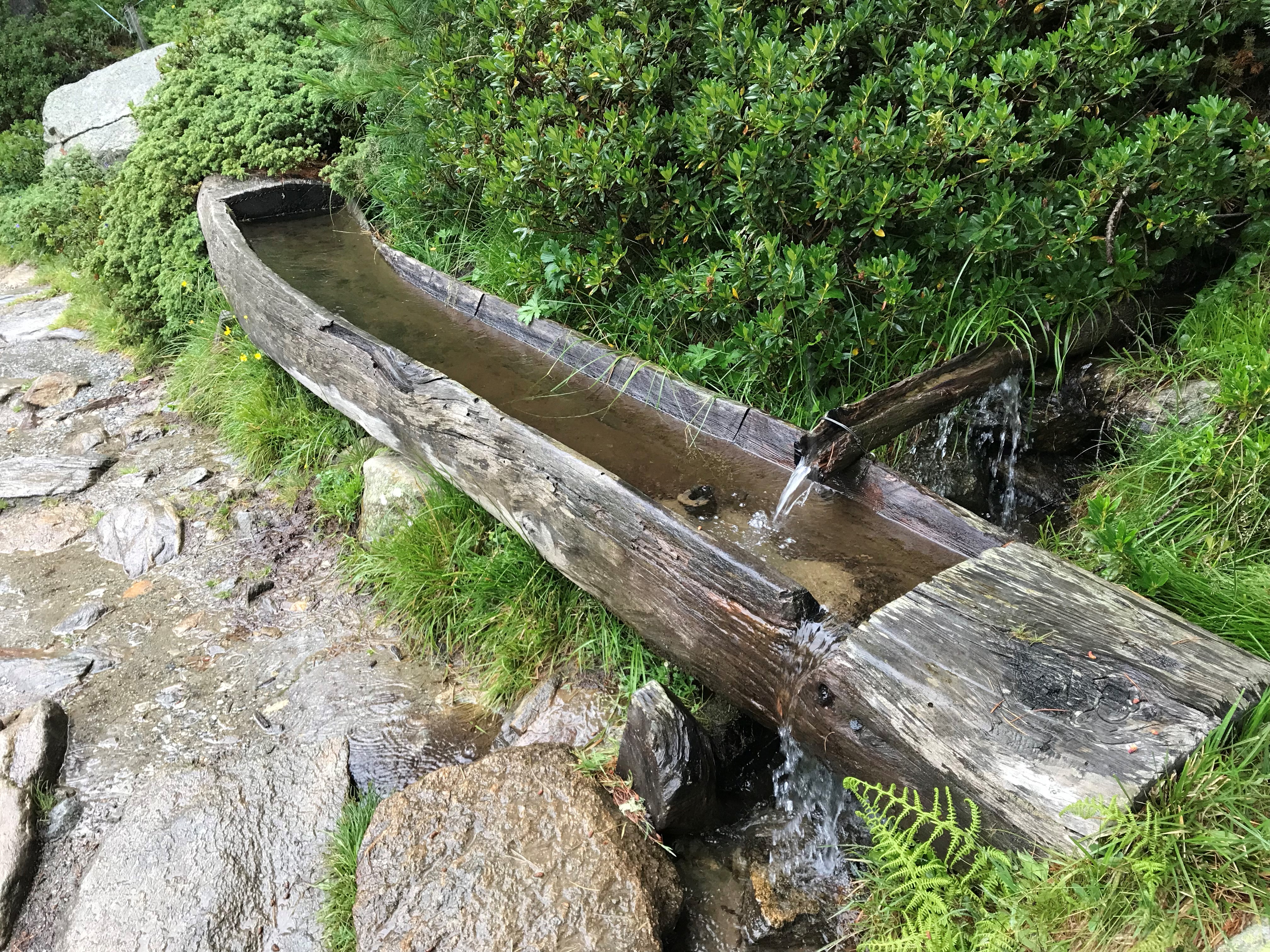}
    \caption{Drinking trough cut in log. Image by \href{https://commons.wikimedia.org/wiki/User:Karelj}{Karelj}, licensed under 
    \href{https://creativecommons.org/licenses/by-sa/4.0/}{CC BY-SA 4.0}, via 
    \href{https://commons.wikimedia.org/wiki/File:Drinking_trough_cut_in_log.jpg}{Wikimedia Commons}.}
    \label{fig:drinking-trough}
\end{figure}

The dynamics of both the fluid flow and the transfer of tracer particles through the control volume depend strongly on its internal characteristics. As an extreme example, consider the drinking trough shown in Fig.\ \ref{fig:drinking-trough}. While the fluid mostly flows directly from the inlet to the outlet, some fluid parcels (and associated tracers) spend significant time in the stagnant part of the control volume. In the field of chemical engineering, the \textit{residence-time distribution} approach was developed to describe the fluid behaviour in a given control volume. The residence-time distribution $E(t)$ (1/s) is defined as \cite{Danckwerts1953,Rodrigues2021}:
\begin{equation}
    E(t) = \frac{C_{\textit{out}}(t)}{\int_{0}^{\infty}C_{\textit{out}}(t')dt'},
\end{equation}
under the condition of pulse tracer insertion at the inlet, i.e.\ $C_{\textit{in}}(t) = \delta(0)$, where $\delta$ is the Dirac function. Because the tracer is considered diluted and does not affect the hydraulics, the flow system is linear and time-invariant with respect to the tracer. In control-systems language, therefore, $E(t)$ is the impulse response of the control volume and for an arbitrary inlet concentration evolution $C_{\textit{in}}(t)$, the outlet concentration can be found as the convolution of $C_{\textit{in}}(t)$ and $E(t)$, i.e.:
\begin{equation}
    C_{\textit{out}}(t) = \int_{0}^{\infty}C_{\textit{in}}(t-\tau)E(\tau)d\tau.
    \label{eq:outlet_conc_cst}
\end{equation}
Note that the variable $\tau$ in the above integral is commonly referred to as \textit{age}. By this token, residence-time distribution is also known as the \textit{exit-age distribution}.

For Eq.\ \ref{eq:outlet_conc_cst} to be valid, it is assumed that the tracers in question are neither destroyed nor created in the control volume. However, this is not true in the case of decaying tracers, such as DNPs. Since decay process is a first-order removal reaction with rate $\lambda$, each element that spends an age $\tau$ inside the control volume is attenuated by the survival factor $\exp(-\lambda\tau)$ and, therefore, Eq.\ \ref{eq:outlet_conc_cst} becomes:
\begin{equation}
    C_{\textit{out}}(t) = \int_{0}^{\infty}C_{\textit{in}}(t-\tau)E(\tau)\exp(-\lambda\tau)d\tau.
    \label{eq:outlet_conc_decay}
\end{equation}
This equation captures the dynamics of DNP transport in control volumes without sources or sinks (excluding decay), such as the ex-core region of the reference system shown in Fig.\ \ref{fig:cstr}, where no fission takes place and chemical removal e.g.\ through off-gassing is considered to be neglected. For the ex-core region, the designation $E_e(\tau)$ will be used to represent the ex-core residence-time distribution.

In the core region, precursors are introduced through two mechanisms:
\begin{enumerate}
    \item At the hydraulic inlet to the core with the inflow rate of $QC_{\textit{in}}$. In the reference system shown in Fig.\ \ref{fig:cstr}, this is due to recirculation.
    \item Through fission, i.e.\ internal generation in the volume with a rate taken to be, for the moment, equal to $S$ (units of precursor amount/time). Note that this is inevitably an approximation, as a single scalar production rate cannot capture the spatial aspect of DNP generation. This is in contrast with hydraulic inflow rate, which can be attributed to a well-defined inlet location.
\end{enumerate}
Because the tracer-transport/decay model is linear and time-invariant, the core outlet signal (outflow rate $QC_{\textit{out}}$) can then be written as a sum of two convolution integrals:
\begin{equation}
    QC_{\textit{out}}(t) = \int_{0}^{\infty}QC_{\textit{in}}(t-\tau)E_{c,i}(\tau)\exp(-\lambda\tau)d\tau + \int_{0}^{\infty}S(t-\tau)E_{c,s}(\tau)\exp(-\lambda\tau)d\tau.
    \label{eq:outlet_core}
\end{equation}
Here, $E_{c,i}(\tau)$ is the core inlet $\rightarrow$ outlet residence-time distribution, while $E_{c,s}(\tau)$ is the related \textit{life-expectancy distribution}, i.e.\ probability density that a brand-new tracer born in the core will exit after a further time $\tau$. Given the scalar production rate approximation introduced above, it is expected that it should be possible to relate $E_{c,s}$ to $E_{c,i}$ based on geometric and hydrodynamic arguments.

The age-based framework for DNP transport is conceptually interesting and could be developed further, leveraging the fact that it reduces the dimensionality of advection from three spatial dimensions to one age dimension. This is, however, not the topic of this paper, although some concepts are sketched in the \hyperref[sec:app]{Appendix}. Focussing instead on the primary goal, i.e.\ the derivation of the zero-power transfer function for the system under investigation, it is now possible to take the Laplace transform of Eq.\ \ref{eq:outlet_conc_decay} with $E(\tau)=E_e(\tau)$ and of Eq.\ \ref{eq:outlet_core}. Using the theorems for the Laplace transform of a convolution and the shifting property, it is obtained:
\begin{align}
    C_{\textit{out},e} &= C_{\textit{in},e}(s)E_e(s+\lambda),\\
    QC_{\textit{out},c} &= QC_{\textit{in},c}(s)E_{c,i}(s+\lambda) + S(s)E_{c,s}(s+\lambda).
\end{align}
Note that, for brevity, the same symbols are used in this work for functions and dependent variables in the time-domain and their respective Laplace transforms. The difference is indicated by the independent variable, time $t$ in the time domain or complex frequency $s$ (1/s) in the frequency domain. Considering now the reference system in Fig.\ \ref{fig:cstr}, $C_{\textit{out},e}=C_{\textit{in},c}$ and $C_{\textit{out},c}=C_{\textit{in},e}$. As a result, it can be deduced that:
\begin{equation}
    QC_{\textit{in},c}(s) = \frac{E_{c,s}(s+\lambda)E_e(s+\lambda)S(s)}{1-E_{c,i}(s+\lambda)E_e(s+\lambda)}.
    \label{eq:cl_qc}
\end{equation}
This equation gives the relation between the concentration of DNPs recirculating into the core and the source rate in-core due to fission. To complete the system, the point kinetics equations will be used in the next section.

\subsection{Coupling with point kinetics}
So far, aside from the assumption of destruction of precursors via decay, no particular aspect of nuclear reactor dynamics has been introduced into the reasoning. If the point-kinetics power equation without external neutron source is cast as \cite{Akcasu1971}:
\begin{equation}
	\frac{dP}{dt} = \frac{\beta}{\Lambda}[k(t)-1]P(t) + \sum_j\lambda_j\Omega_j(t),
	\label{eq:power_pk}
\end{equation}
then $P(t)$ is the fission rate (1/s), $\Lambda$ (s) the prompt neutron generation time, and $\Omega_j$ the importance-weighted integral of concentration of precursors in group $j$ over the active core (1/s). In this formulation, the source of precursors in group $j$ due to fission is:
\begin{equation}
	S_j(t) = \frac{\beta_j}{\Lambda}P(t),
	\label{eq:sj}
\end{equation}
with units 1/s$^2$. The units of $C$ as defined in the previous sections become 1/(s$\cdot$m$^3$) due to the selected normalisation. 

For notation simplification purposes, only one group of precursors will be considered in the following derivation with the extension to multiple groups following naturally. Furthermore, the difference between the core residence-time distribution $E_{c,i}$ and the life-expectancy distribution $E_{c,s}$ will be neglected for simplicity (i.e.\ $E_{c,s} = E_{c,i} \equiv E_c$) and it will be re-introduced after the derivation is completed. Still further, importance weighting of precursors will be ignored for simplicity in this work, assuming zero importance ex-core and uniform importance in-core. Depending on the reactor design and flow conditions, this assumption can have a significant impact on the results. For example, in the extreme case of infinite flow velocity redistributing the precursors throughout the active core uniformly, it can be easily imagined that importance weighting strongly affects the value of $\Omega_j$. A more quantitative discussion of this topic can be found e.g.\ in \cite{Aufiero2014}. Thus, neglecting of importance weighting is a deficiency impairing the generality of the presented theoretical framework, which should be remedied in future work. For the time being, some thoughts on the inclusion of spatial importance in an age-based DNP transport framework are given in the \hyperref[sec:app]{Appendix}.

With the above simplifications, $\Omega(t)$ becomes equal to the precursor ``burden'' present in the active core at time $t$. This is equal to \cite{Schwartz1979}:
\begin{equation}
	\Omega(t) = \int^{\infty}_{0}A(t,t-\tau)d\tau.
    \label{eq:omega}
\end{equation}
Here, $A(t,t_0)$ is the persistence function, i.e.\ the amount of precursors introduced to the core per unit time at time $t_0$ remaining there at time $t$ \cite{Schwartz1979}. To derive the persistence function, one needs to consider the rate of introduction and the effects of both decay and outflow. If the rate of introduction is labelled as $F$, then:
\begin{equation}
	A(t,t-\tau) = F(t-\tau)\exp(-\lambda\tau) - \int_{0}^{\tau}F(t-\tau)E_c(t')\exp(-\lambda t')\cdot\exp[-\lambda(\tau-t')]dt'
	\label{eq:persistence}
\end{equation}
The first part of the expression incorporates the effect of decay and the second one the effect of outflow. In Eq.\ \ref{eq:persistence}, the outflow expression based on residence-time distribution (see Eq.\ \ref{eq:outlet_core}) was used. Note the two decay terms multiplied together -- the first one corresponds to the decay of precursors before they exit the core, while the second one is introduced to satisfy mass conservation at time $t$. The need for this can be seen more clearly if one considers the special case $E_c(t') = H(\tau_c-t')/\tau_c$, where $H(t)$ is the Heaviside step function. In this case, $A(t,t-\tau_c)$ must be equal to zero as all precursors have at that point either decayed or exited the core via outflow. Introducing this special case to Eq.\ \ref{eq:persistence}, it can be seen that this condition will be satisfied as:
\begin{align}
	\begin{split}
		\int_{0}^{\tau_c}F(t-\tau_c)E_c(t')\exp(-\lambda \tau_c)dt' &= \frac{1}{\tau_c}F(t-\tau_c)\exp(-\lambda \tau_c)\int_{0}^{\tau_c} H(\tau_c-t')dt' \\&= \frac{1}{\tau_c}F(t-\tau_c)\exp(-\lambda \tau_c)\int_{0}^{\tau_c}dt' = F(t-\tau_c)\exp(-\lambda \tau_c).
	\end{split}
\end{align}
Coming back to Eq.\ \ref{eq:persistence}, this can now be simplified as: 
\begin{equation}
		A(t,t-\tau) = F(t-\tau)\exp(-\lambda\tau) \bigg(1-\int_{0}^{\tau}E_c(t')dt'\bigg).
        \label{eq:persistence_2}
\end{equation}
It is noted in passing that one can define the core \textit{internal-age distribution} $I_c(\tau)$ (1/s), i.e.\ function such that $I_c(\tau)d\tau$ is the fraction of fluid elements inside the core with age between $\tau$ and $\tau + d\tau$ \cite{Rodrigues2021}. It can be shown that the following relation holds \cite{Bolin1973,Rodrigues2021}:
\begin{equation}
    I_c(\tau) = \frac{1}{\tau_c}\bigg(1-\int_{0}^{\tau}E_c(t')dt'\bigg).
\end{equation}
Here, $\tau_c$ is the first moment of the residence-time distribution, i.e.\ the mean residence time. With this definition, $\Omega(t)$ can be written as:
\begin{equation}
	\Omega(t) = \tau_c\int_{0}^{\infty}F(t-\tau)\exp(-\lambda\tau) I_c(\tau)d\tau.
	\label{eq:omega_t}
\end{equation}

Coming back to Eq.\ \ref{eq:omega} with the persistence function given according to Eq.\ \ref{eq:persistence_2}, $\Omega(t)$ is written as a convolution \cite{Rodrigues2021}:
\begin{equation}
    \Omega(t) = F(t) * \bigg[\exp(-\lambda t) \bigg(1-\int_{0}^{t}E_c(t')dt'\bigg)\bigg]
\end{equation}
Using the theorems for Laplace transform of a convolution, linearity, and the shifting property leads to:
\begin{equation}
	\Omega(s) = F(s)\cdot\frac{1-E_c(s+\lambda)}{s+\lambda}.
\end{equation}
Now, extending this result by considering individually $F_i(s) = QC_\textit{in}(s)$ and $F_s(s) = S(s)$ with $E_{c,i} \neq E_{c,s}$:
\begin{equation}
	\Omega(s) = QC_\textit{in}(s)\cdot\frac{1-E_{c,i}(s+\lambda)}{s+\lambda} + S(s)\cdot\frac{1-E_{c,s}(s+\lambda)}{s+\lambda}.
\end{equation}
Substituting from Eq.\ \ref{eq:cl_qc}:
\begin{align}
	\begin{split}
		\Omega(s) &= \bigg[\frac{E_{c,s}(s+\lambda)E_e(s+\lambda)}{1-E_{c,i}(s+\lambda)E_e(s+\lambda)}\frac{1-E_{c,i}(s+\lambda)}{s+\lambda} + \frac{1-E_{c,s}(s+\lambda)}{s+\lambda}\bigg]S(s) \\
		&= \frac{1}{s+\lambda}\frac{1-E_{c,s}(s+\lambda)+E_e(s+\lambda)[E_{c,s}(s+\lambda)-E_{c,i}(s+\lambda)]}{1-E_{c,i}(s+\lambda)E_e(s+\lambda)}\cdot S(s).
        \label{eq:omega_s}
	\end{split}
\end{align}
With this equation, the zero-power transfer function and static loss of reactivity due to DNP drift can now be derived.

\subsection{Zero-power transfer function}
\label{sec:zeropower}
In order to derive the zero-power transfer function $Z(s)$, Eq.\ \ref{eq:power_pk} must be first linearised around the steady state with $P(t) = P_0 + p(t)$, $k(t) = k_0 + k_1(t)$, $\Omega_k(t) = \Omega_{j,0} + \omega_j(t)$ as:
\begin{equation}
	\frac{dp}{dt} = \frac{\beta}{\Lambda}[k_1(t) P_0 + (k_0-1)p(t)] + \sum_j\lambda_j\omega_j(t).
\end{equation}
In the frequency domain, this leads to:
\begin{equation}
	sp(s) = \frac{\beta}{\Lambda}[k_1(s) P_0 + (k_0-1)p(s)] + \sum_j\lambda_j\omega_j(s)
\end{equation}
and after re-arranging and multiplying by $\Lambda/\beta/P_0$:
\begin{equation}
	\bigg[\frac{\Lambda}{\beta}s-(k_0-1)\bigg]\frac{p(s)}{P_0} = k_1(s) + \sum_j\lambda_j\frac{\Lambda}{\beta}\frac{\omega_j(s)}{P_0}.
	\label{eq:intermediate_z}
\end{equation}
From Eqs.\ \ref{eq:sj} and \ref{eq:omega_s} it is known that:
\begin{equation}
	\omega_j(s) = \frac{1}{s+\lambda_j}\frac{1-E_{c,s}(s+\lambda_j)+E_e(s+\lambda_j)[E_{c,s}(s+\lambda_j)-E_{c,i}(s+\lambda_j)]}{1-E_{c,i}(s+\lambda_j)E_e(s+\lambda_j)}\cdot\frac{\beta_j}{\Lambda}p(s).
\end{equation}
Substituting this into Eq.\ \ref{eq:intermediate_z} and re-arranging gives the zero-power transfer function $Z(s) = p(s)/P_0/k_1(s)$:
\begin{align}
	\begin{split}
	Z(s) = \bigg[&\frac{\Lambda}{\beta}s-(k_0-1) \\&-\sum_j\frac{\lambda_j}{s+\lambda_j}\frac{1-E_{c,s}(s+\lambda_j)+E_e(s+\lambda_j)[E_{c,s}(s+\lambda_j)-E_{c,i}(s+\lambda_j)]}{1-E_{c,i}(s+\lambda_j)E_e(s+\lambda_j)}\cdot\frac{\beta_j}{\beta}\bigg]^{-1}.
	\end{split}
	\label{eq:Z}
\end{align}
Since Eq.\ \ref{eq:omega_s} for $s=0$ gives the steady-state precursor burden in the core, it is straightforward to use Eq.\ \ref{eq:power_pk} in its steady-state form to derive, in a similar manner, the reactivity loss due to DNP drift as:
\begin{equation}
	k_0 = 1-\sum_j\frac{1-E_{c,s}(\lambda_j)+E_e(\lambda_j)[E_{c,s}(\lambda_j)-E_{c,i}(\lambda_j)]}{1-E_{c,i}(\lambda_j)E_e(\lambda_j)}\cdot\frac{\beta_j}{\beta}.
    \label{eq:k0}
\end{equation}
The reader is reminded that in Eq.\ \ref{eq:k0}, the evaluation of the distributions is done in the frequency domain for $s=\lambda_j$. This concludes the derivations in this work and the results can now be used for a quantitative study of the system shown in Fig.\ \ref{fig:cstr}.

As a side note, Eq.\ \ref{eq:Z} can be slightly manipulated as:
\begin{align}
	\begin{split}
	Z(s) = \bigg[&\frac{\Lambda}{\beta}s-k_0+\bigg(1-\sum_j\frac{\lambda_j}{s+\lambda_j}\cdot\\&\cdot\frac{1-E_{c,s}(s+\lambda_j)+E_e(s+\lambda_j)[E_{c,s}(s+\lambda_j)-E_{c,i}(s+\lambda_j)]}{1-E_{c,i}(s+\lambda_j)E_e(s+\lambda_j)}\cdot\frac{\beta_j}{\beta}\bigg)\bigg]^{-1}.
	\end{split}
\end{align}
Defining then $k_p(s)$ as:
\begin{equation}
	k_p(s) = 1-\sum_j\frac{\lambda_j}{s+\lambda_j}\frac{1-E_{c,s}(s+\lambda_j)+E_e(s+\lambda_j)[E_{c,s}(s+\lambda_j)-E_{c,i}(s+\lambda_j)]}{1-E_{c,i}(s+\lambda_j)E_e(s+\lambda_j)}\cdot\frac{\beta_j}{\beta}
\end{equation}
leads to the compact formulation:
\begin{equation}
	Z(s) = \bigg[\frac{\Lambda}{\beta}s+k_p(s)-k_p(0)\bigg]^{-1},
\end{equation}
since $k_0 = k_p(0)$.

\subsection{Gamma distribution}
\label{sec:gamma}
\begin{figure}
	\centering
	\includegraphics[width=1\linewidth]{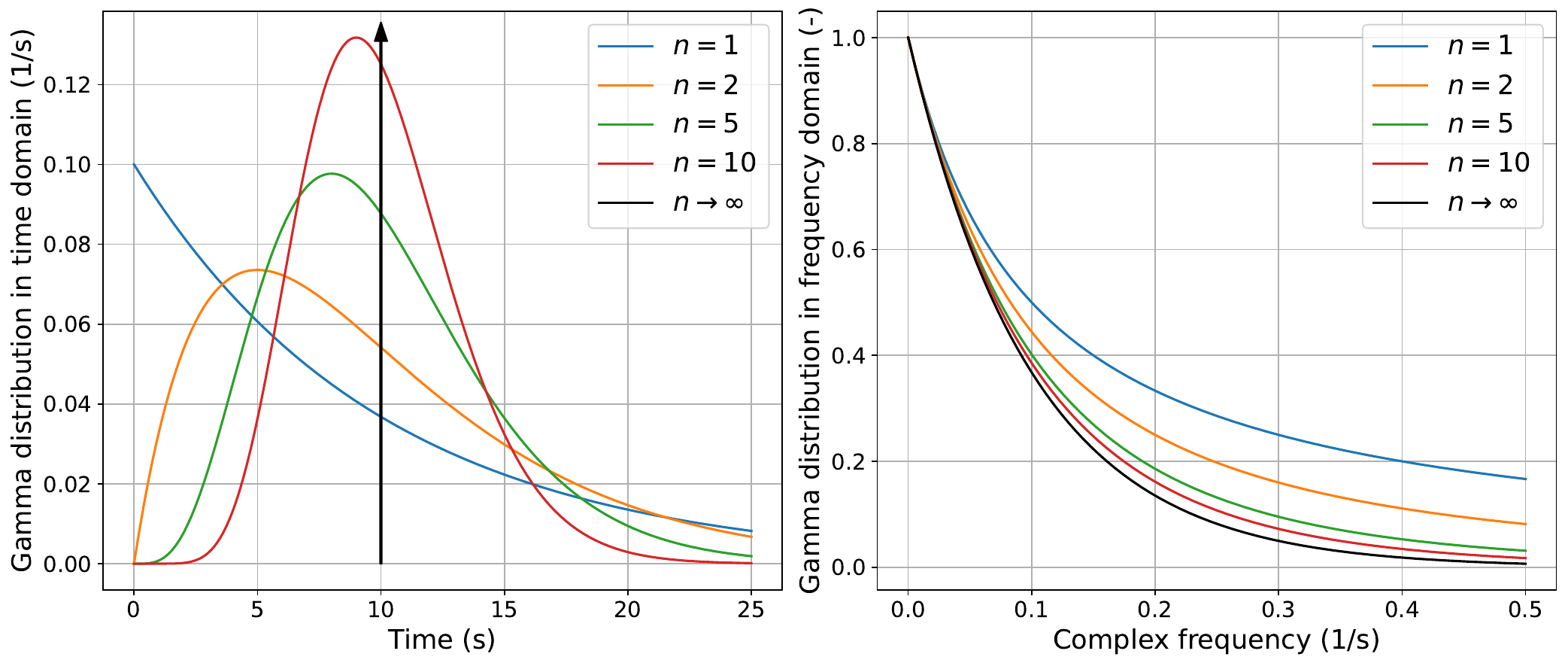}
	\caption{Gamma distribution in the time domain (Eq.\ \ref{eq:gamma_time}, left) and in the frequency domain (Eq.\ \ref{eq:gamma_time}, right) for $\tau_m=10\ \text{s}$ and several representative values of $n$.}
	\label{fig:gamma}
\end{figure}

In the previous sections, no restrictions were placed on the residence-time distributions, leaving them general. In practice, a certain shape of the distribution must be chosen. A typical choice from the field of chemical engineering is the gamma distribution, which can be cast in the time domain as \cite{MacMullin1935,Toson2019}:
\begin{equation}
	E(t;n,\tau_m) = \frac{(n/\tau_m)^n}{\Gamma(n)}t^{n-1}e^{-n t/\tau_m}.
	\label{eq:gamma_time}
\end{equation}
Here, $\Gamma(n)$ is the gamma function. Using Eq.\ \ref{eq:gamma_time} to characterise the residence-time distribution has the physical meaning of representing the given region as a system of continuous-flow mixing tanks in series with $\tau_m$ being the first moment of the distribution, i.e.\ the mean residence time, and $n$ can be understood as the number of tanks. The two extreme cases of the gamma distribution are the exponential distribution ($n=1$):
\begin{equation}
    E_1(t;\tau_m) = \frac{1}{\tau_m}e^{-t/\tau_m},
\end{equation}
and the Dirac distribution ($n\rightarrow\infty$):
\begin{equation}
    E_\infty(t;\tau_m) = \delta(t-\tau_m).
\end{equation}
Since the exponential distribution represents a system with perfect mixing and the Dirac distribution one with no mixing in the given region \cite{Rodrigues2021, Bolin1973}, the parameter $n$ can also be understood as the reciprocal of the degree of mixing in the control volume under consideration. Figure \ref{fig:gamma}(left) shows the gamma distribution in time domain for different values of $n$. Note that, for simplicity, configurations with $n<1$ (representing a bypass \cite{MacMullin1935,Toson2019}) are not considered here. It can be seen that the distribution for $n=1$ is qualitatively different from the other values due to the complete absence of a peak. With increasing $n$, the distribution becomes more peaked and symmetric around the mean value $\tau_m$. In the frequency domain:
\begin{equation}
    E(s;n,\tau_m) = \exp\bigg[-n\ln\bigg(1+s\frac{\tau_m}{n}\bigg)\bigg].
    \label{eq:gamma_freq}
\end{equation}
Figure \ref{fig:gamma}(right) shows the Laplace transform of the gamma distribution for different values of $n$. The difference between the shapes of the distributions is less pronounced here than in the time domain. Since Eqs.\ \ref{eq:Z} and \ref{eq:k0} utilise evaluations of the distribution in the frequency domain, rather than in the time domain, it can be expected that for $n\gtrsim 2$ the effect of $n$ on the reactivity loss and frequency response will be limited. Figure \ref{fig:gamma}(right) shows that for $n\gtrsim2$ and $\lambda_j\tau\gtrsim 3$, the value of the distribution is already strongly diminished ($\lesssim 0.1$). Note that for the two extreme cases:
\begin{equation}
    E_1(s; \tau_m) = \frac{1}{1+s\tau_m},
\end{equation}
and:
\begin{equation}
    E_\infty(s; \tau_m)= \exp(-s\tau_m),
\end{equation} 

It is remarked that this framework is a generalisation of some of the previously-used simplified models of CFR dynamics. As a final derivation, it is shown here how it can be used to derive such an existing model, the ``CSTR model'' (Continuous-Stirred Tank Reactor model). In the present formalism, the CSTR model of point kinetics refers to a configuration where the core residence-time distribution is the exponential distribution ($n=1$) with mean residence time $\tau_c$, $E_{c,s}\equiv E_{c,i}$, and the ex-core residence-time distribution is the Dirac distribution ($n\rightarrow\infty$) with mean residence time $\tau_e$, see e.g.\ \cite{Cammi2010, Bures2024b}. Then, Eq.\ \ref{eq:omega_s} becomes:
\begin{align}
	\begin{split}
		\Omega(s) &= \frac{1}{s+\lambda}\cdot\frac{1-\big[1+(s+\lambda)\tau_c\big]^{-1}}{1-\big[1+(s+\lambda)\tau_c\big]^{-1}\exp\big[-(s+\lambda)\tau_e\big]}\cdot S(s)\\
				  &= \frac{1}{s+\lambda}\cdot\frac{(s+\lambda)\tau_c}{(s+\lambda)\tau_c+1-\exp\big[-(s+\lambda)\tau_e\big]}\cdot S(s)\\
				  &= \frac{1}{s+\lambda + \Big(1-\exp\big[-(s+\lambda)\tau_e\big]\Big)\Big/\tau_c}\cdot S(s).
	\end{split}
\end{align}
Introducing this into the derivations in Section \ref{sec:zeropower} leads to:
\begin{equation}
	Z(s) = \bigg[\frac{\Lambda}{\beta}s-(k_0-1)-\sum_j\frac{\lambda_j}{s+\lambda_j + \Big(1-\exp\big[-(s+\lambda_j)\tau_e\big]\Big)\Big/\tau_c}\cdot\frac{\beta_j}{\beta}\bigg]^{-1},
	\label{eq:cstr_z}
\end{equation}
and:
\begin{equation}
	k_0 = 1-\sum_j\frac{\lambda_j}{\lambda_j + \Big(1-\exp\big[-\lambda_j\tau_e\big]\Big)\Big/\tau_c}\cdot\frac{\beta_j}{\beta}.
	\label{eq:cstr}
	\end{equation}
This is equivalent to the conventionally reported solution \cite{Cammi2010, Bures2024b}.

\section{Quantitative analysis}
\label{sec:res}

For the purpose of this work, if not specified otherwise, $E_{c,s}$, $E_{c,i}$, and $E_e$ are taken to be gamma distributions parameterised by $n_{c,s}=n_{c,i}\equiv n_c$, $\tau_{c,i}\equiv\tau_c$, $\tau_{c,s}=\tau_c/2$, and $n_e$ and $\tau_e$. In other words, the degree of mixing in the core is taken to be the same for DNPs born from fission and those arriving at the inlet, i.e.\ $n_c$ is taken to be a property of the core region. For example, a channel-type reactor with a well-defined flow path from the inlet to the outlet such as the Molten Salt Reactor Experiment (MSRE) could be represented by a Dirac distribution with $n_c\rightarrow\infty$ and a reactor with complex flow patterns such as the Molten Salt Fast Reactor (MSFR) \cite{Martin2024} could be anticipated to have $n_c$ much closer to 1. Additionally, the mean residence time of DNPs born from fission is taken to be half of the one for those arriving at the inlet. This approximation attempts to capture the longer physical distance that must be travelled by particles from the inlet to the outlet in comparison to those created within the core region.

\subsection{Reactivity loss}
Equation \ref{eq:k0} can be simplified as:
\begin{equation}
	k_0 = \sum_j k_{0,j} = \sum_j\frac{[1-E_e(\lambda_j)]E_{c,s}(\lambda_j)}{1-E_{c,i}(\lambda_j)E_e(\lambda_j)}\cdot\frac{\beta_j}{\beta},
	\label{eq:k0_1}
\end{equation}
This is one of the figures of merit that will be used in the analysis. Furthermore, for an individual DNP group, the reactivity loss scaled by the DNP fraction is defined as:
\begin{equation}
		\kappa(\lambda_j) = \frac{k_{0,j}}{\beta_j/\beta} =\frac{[1-E_e(\lambda_j)]E_{c,s}(\lambda_j)}{1-E_{c,i}(\lambda_j)E_e(\lambda_j)}.
		\label{eq:scaled_loss}
\end{equation}
For analysing the effect of recirculation, it is noted that a system without recirculation would have $\tilde{E}_e(s)\equiv 0$ with Eq.\ \ref{eq:k0_1} becoming:
\begin{equation}
    \tilde{k}_0 = \sum_j E_{c,s}(\lambda_j)\cdot\frac{\beta_j}{\beta}.
\end{equation}
The overall relative recirculation reactivity effect can then be defined as:
\begin{equation}
    \varepsilon = \frac{\tilde{k}_0-k_0}{\tilde{k}_0}.
    \label{eq:recirc_effect_all}
\end{equation}
For an individual DNP group, it is taken as:
\begin{equation}
    \varepsilon(\lambda_j) = \frac{\tilde{k}_{0,j}-k_{0,j}}{\tilde{k}_{0,j}} = \frac{[1-E_{c,i}(\lambda_j)]E_e(\lambda_j)}{1-E_{c,i}(\lambda_j)E_e(\lambda_j)}.
    \label{eq:recirc_effect}
\end{equation}

\begin{figure}
	\makebox[\textwidth][c]{%
		\includegraphics[width=1.1\linewidth]{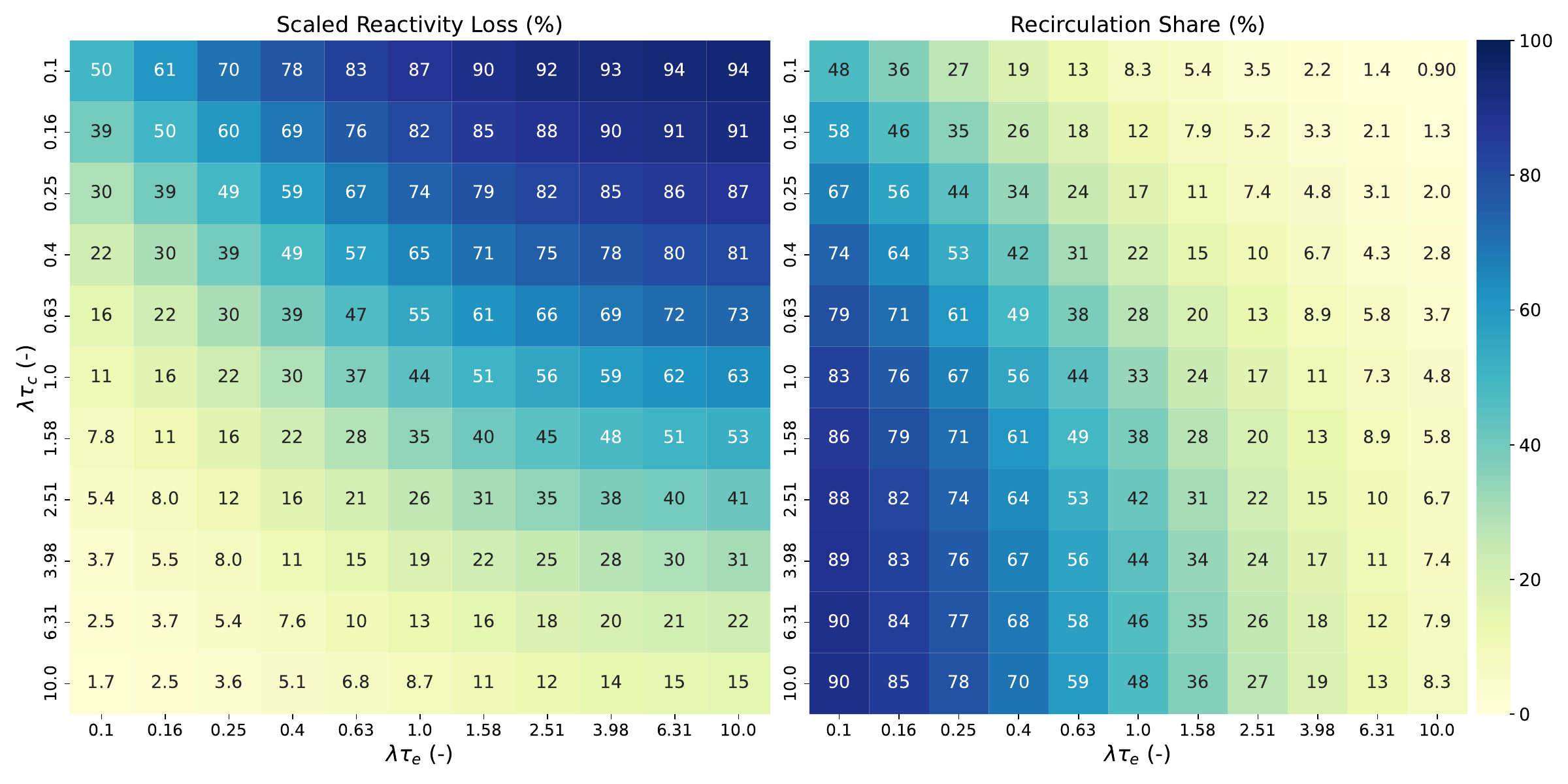}
	}
	\caption{Scaled reactivity loss (Eq.\ \ref{eq:scaled_loss}, left) and relative recirculation effect (Eq.\ \ref{eq:recirc_effect}, right) as functions of $\lambda\tau_c$ and $\lambda\tau_e$ for a moderate level of mixing ($n=3$) both in-core and ex-core. Left: 100\% = complete loss of DNPs. Right: 100\% = all DNPs leaving the core recirculate back Logarithmic scaling is used for the axes in both figures.}
	\label{fig:heatmapfigure}
\end{figure}

A key observation derived from the presented theory is that in Eq.\ \ref{eq:gamma_freq}, $E$ depends only on the product of $s\tau$, rather on the values of these two parameters individually. Therefore, for individual DNP groups, it is possible to investigate drift as a function of $\lambda\tau$, reducing the parameter space. In the reference problem considered, the two governing parameters are $\lambda\tau_c$ and $\lambda\tau_e$, i.e.\ products of the residence time in the two control volumes and precursor decay constants. To understand the physical significance of these variables, one can take for example the limit $\lambda\tau_c\rightarrow 0$: this corresponds to either negligible decay or negligible in-core residence time. Conversely, $\lambda\tau_c\rightarrow \infty$ implies immediate decay or in-core residence time approaching the one of static fuel. In other words, they could be understood as ``optical residence times'' of the two control volumes for precursors of a given decay constant $\lambda$. Figure \ref{fig:heatmapfigure} shows the scaled reactivity loss (Eq.\ \ref{eq:scaled_loss}) and relative DNP recirculation contribution to reactivity (Eq.\ \ref{eq:recirc_effect}) as functions of $\lambda\tau_c$ and $\lambda\tau_e$ for a moderate level of mixing ($n=3$) both in-core and ex-core. Varying $n_c$ and/or $n_e$ does not significantly alter the qualitative pattern of the data; it only affects the slopes with respect to $\lambda\tau_c$ and $\lambda\tau_e$.

Looking at the results in Fig.\ \ref{fig:heatmapfigure}, the following can be discerned:
\begin{enumerate}
	\item As expected, reactivity loss due to precursor drift increases with $\lambda\tau_e$ and decreases with $\lambda\tau_c$. Thus, systems with ex-core residence time which is comparatively low with respect to the in-core residence time exhibit a low reactivity loss. However, the effect of $\lambda\tau_c$ is more pronounced than the one of $\lambda\tau_e$, i.e.\ holding $\tau_c$ and $\tau_e$ constant and increasing $\lambda$ (i.e.\ decreasing DNP half-life) should decrease the loss due to drift.
	\item Conversely, recirculation fraction decreases with $\lambda\tau_e$ and increases with $\lambda\tau_c$. That means systems with ex-core residence time which is comparatively low with respect to the in-core residence time exhibit a high relative role of recirculated DNPs on reactivity. The effect of $\lambda\tau_e$ is more pronounced than the one of $\lambda\tau_c$; this implies that holding $\tau_c$ and $\tau_e$ constant and increasing $\lambda$ will decrease the recirculation fraction.
\end{enumerate}
The combination of these effects implies that, for example, for low $\lambda\tau_e$ and high $\lambda\tau_c$, the DNP reactivity recirculation fraction is significant but the total reactivity loss is small, reducing the relevance of the DNP advection effect overall. The most ``interesting'' dynamics can be expected when both the reactivity loss and recirculation fraction are appreciable -- this occurs when $\tau_e\approx\tau_c$ and $\lambda$ is not significantly large with respect to $1/\tau_c$.

\begin{figure}
	\centering
	\includegraphics[width=1\linewidth]{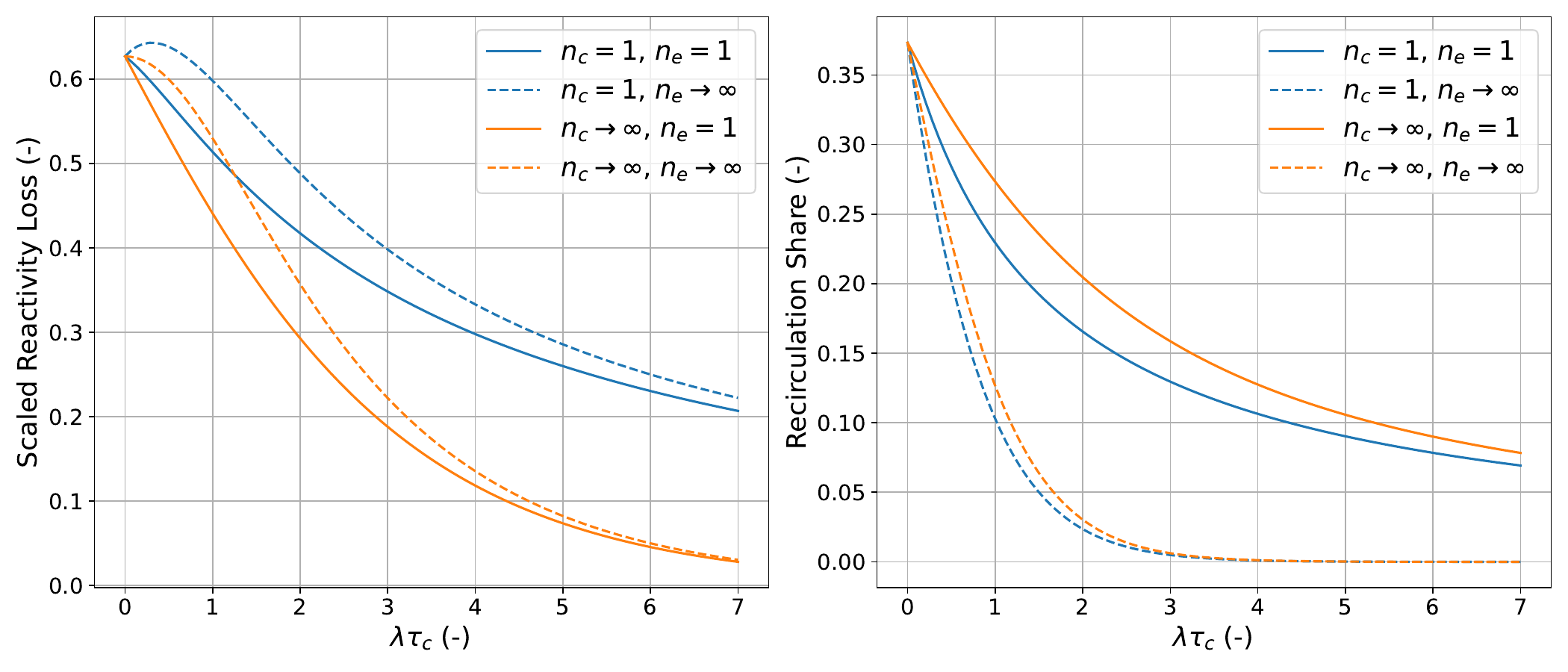}
	\caption{Scaled reactivity loss (Eq.\ \ref{eq:scaled_loss}, left) and relative recirculation effect (Eq.\ \ref{eq:recirc_effect}, right) as functions of $\lambda\tau_c$ with $\tau_e/\tau_c=1.68$ (MSRE-like) for bounding cases of mixing in-core and ex-core. Left: 1 = complete loss of DNPs. Right: 1 = all DNPs leaving the core recirculate back.}
	\label{fig:msretaulambda}
\end{figure}

Currently, the main class of CFRs under investigation are liquid-fuelled molten-salt reactors. A review of their designs reveals that $\tau_e\approx\tau_c$ is a rather common occurrence, for example:
\begin{itemize}
	\item Molten Salt Reactor Experiment (MSRE): $\tau_c=9.4$ s, $\tau_e=15.8$ s \cite{Fratoni2020},
	\item Denatured Molten Salt Reactor: $\tau_c=15$ s, $\tau_e=11$ s \cite{Engel9180},
	\item Reference Molten Salt Fast Reactor (MSFR): $\tau_c=2$ s, $\tau_e=2$ s \cite{Brovchenko2019}.
\end{itemize}
Taking the MSRE-like $\tau_e/\tau_c=1.68$ ratio, the resulting scaled reactivity loss (Eq.\ \ref{eq:scaled_loss}) and relative DNP recirculation contribution to reactivity (Eq.\ \ref{eq:recirc_effect}) as functions of $\lambda\tau_c$ are shown in Fig.\ \ref{fig:msretaulambda} for the bounding values of $n_c$ and $n_e$. One can see that, independently of the model, in the limit $\lambda\tau_c\rightarrow0$ the reactivity loss converges to $\tau_e/\tau_t\approx0.63$ with $\tau_t = \tau_c + \tau_e$ being the total system residence time, as expected. It can also be discerned that the reactivity loss is strongly affected by the degree of mixing in-core, while the recirculation share depends rather on the degree of mixing ex-core.

In general, it can be seen that both reactivity loss and recirculation share decrease with $\lambda$ and increase with flow rate (which has the effect of simultaneously diminishing $\tau_c$ and $\tau_e$) but that recirculation contribution drops more rapidly. This is in line with the conclusions drawn from the data in Fig.\ \ref{fig:heatmapfigure}. However, it is also possible to observe that for $n_c=1$ and $n_e\rightarrow\infty$ the evolution of the reactivity loss is non-monotonic in the vicinity of $\lambda\tau_c\rightarrow0$. This peculiar phenomenon can be explained by returning to the case $\tau_{c,i}\neq\tau_{c,s}\neq\tau_e$ (augmenting the definition $\tau_t = \tau_{c,i} + \tau_e$). Coming back to Eq.\ \ref{eq:scaled_loss}, the numerator can be understood as the amount of DNPs born from fission entering the ex-core and decaying there $(1-E_e)E_{c,s}$ multiplied by the recirculation effect $1/(1-E_e E_{c,i})$. Increasing $\lambda$ increases decay of DNPs both in-core and ex-core. If the DNP source is located ``closer'' (in terms of mean in-core residence time) to the exit of the core, it is conceivable that the impact on decay ex-core is more significant than on the decay in-core and, therefore, precursor drift can increase with $\lambda$ under certain conditions. To analyse this in a tractable manner, one can concentrate on the simple no-mixing case with $n_c$ and $n_e$ going to infinity. Then, Eq.\ \ref{eq:scaled_loss} becomes:
\begin{equation}
	\kappa(\lambda) = \exp(-K_s\lambda \tau_e)\frac{1-\exp(-\lambda\tau_e)}{1-\exp(-K_t\lambda\tau_e)}.
\end{equation}
Here, $K_t=\tau_t/\tau_e$ and $K_s=\tau_{c,s}/\tau_e$. Expanding this expression around $x=\lambda\tau_e=0$ and simplifying gives:
\begin{equation}
	\kappa(\lambda) = \frac{\tau_e}{\tau_t} + \lambda\cdot\bigg(1 - 2\frac{\tau_{c,s}}{\tau_{c,i}}\bigg)\cdot\frac{\tau_{c,i}\tau_e}{2\tau_t} + \mathcal{O}(x^2).
	\label{eq:taylor}
\end{equation}
According to this equation, the requirement for the ex-core decay being impacted more by an increase of $\lambda$ than the in-core one is given as $\tau_{c,s}<\tau_{c,i}/2$. More generally in the case of gamma residence-time distributions, the exact form of this expression can be shown to be strongly dependent on the degree of mixing assumed in-core and ex-core. Nevertheless, the necessary condition is that the in-core mean residence time of DNPs born from fission $\tau_{c,s}$ is less than the one for recirculating DNPs $\tau_{c,i}$ by some factor dependent on $n_c$ and $n_e$. Note that Eq.\ \ref{eq:taylor} also implies that, if this is satisfied, an increase in flow rate \textit{decreases} static reactivity loss, since all values of $\tau$ decrease simultaneously with flow rate. It should be remarked that for this phenomenon recirculation is crucial, since without recirculation:
\begin{equation}
	\tilde{\kappa}(\lambda) = \exp(-\lambda \tau_s) = 1-\lambda\tau_s + \mathcal{O}(x^2),
\end{equation}
and reactivity loss only decreases with $\lambda$ and $\tau$.

\begin{table}
	\centering
	\caption{Precursor group data for the $^{235}$U loading of the MSRE, $\beta=666.1$ pcm \cite{Prince1968}.}
	\begin{tabular}{c|cccccc}
		\hline
		Group & 1 & 2 &3 & 4 & 5 & 6  \\ \hline\rule{0pt}{2ex}
		$\lambda_j$ (1/s) & 0.0124 & 0.0305 & 0.1114 & 0.3013 & 1.140 & 3.010 \\ 
		$\beta_j$ (pcm) & 22.3 & 145.7 & 130.7 & 262.8 & 76.6 & 28.0 \\ 
		$\beta_j/\beta$ (\%) & 3.35 & 21.87 & 19.62 & 39.45 & 11.50 & 4.20 \\ \hline
	\end{tabular}
	\label{tab:msrednp}
\end{table}

\begin{figure}
\centering
\includegraphics[width=1\linewidth]{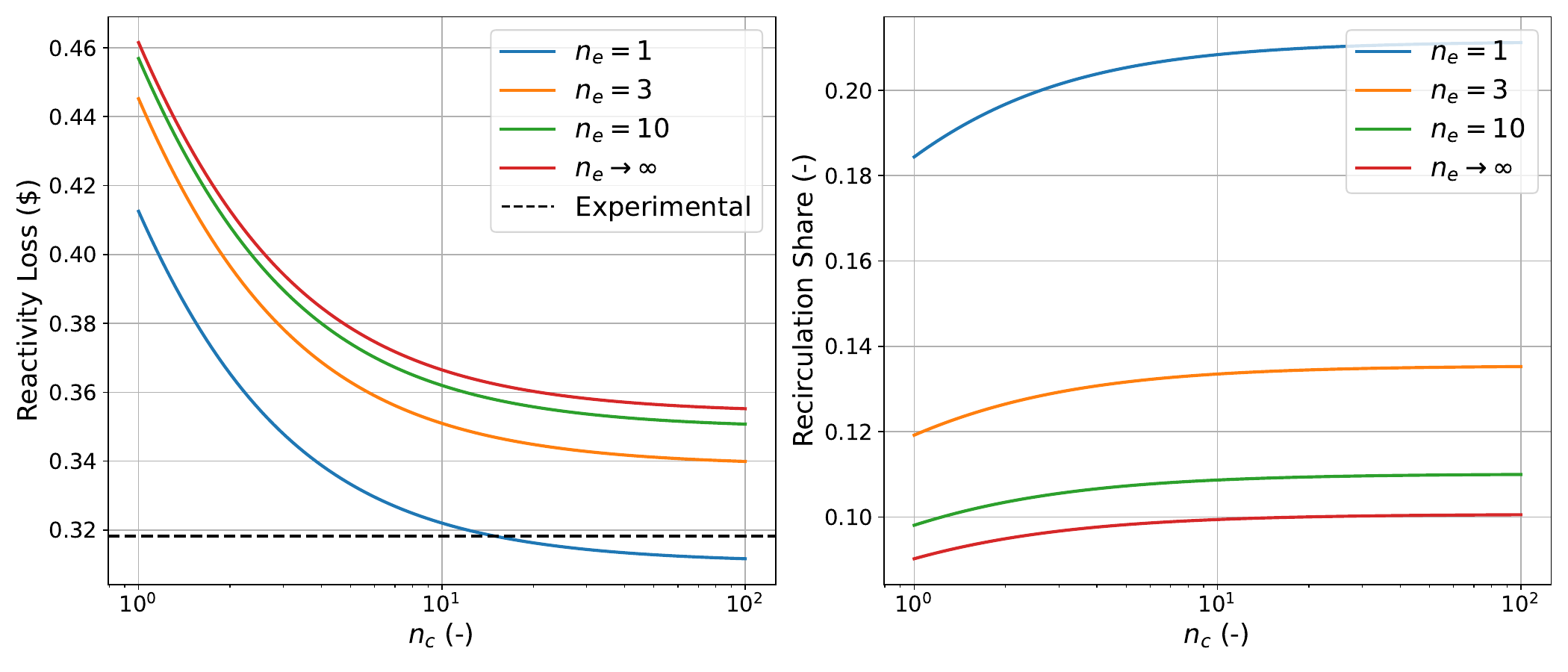}
\caption{Static reactivity loss (Eq.\ \ref{eq:k0_1}, left) and relative recirculation effect (Eq.\ \ref{eq:recirc_effect_all}, right) as functions of $n_c$ for selected values of $n_e$ in an MSRE-like system. The experimental result $k_0=0.318$ \cite{Prince1968} is also shown on the left.}
\label{fig:msrenc}
\end{figure}

To conclude the analysis of precursor drift, two benchmark problems were analysed. In the first one, the static reactivity loss and DNP recirculation contribution to reactivity were investigated for an MSRE-like system, i.e.\ $\tau_c=9.4$ s, $\tau_e=15.8$ s and six precursor groups as per Table \ref{tab:msrednp}, with the results given in Figure \ref{fig:msrenc}. Looking at the static reactivity loss specifically, it appears that in order to match the experimental result, a low degree of mixing (high $n$) should be considered in-core and high degree of mixing (low $n$) ex-core. This is in-line with the fact that the MSRE had a channel-type reactor core with predominantly laminar flow regime \cite{Engel1962}, for which low mixing can be expected. Conversely, majority of the ex-core region was composed of large volumes \cite{Fratoni2020}, such as the lower and upper reactor plenum, for which mixing can be anticipated to be significant. This finding is in direct contradiction to the conventional Continuous-Stirred Tank Reactor model (Eq.\ \ref{eq:cstr}), for which no mixing is assumed ex-core and complete mixing in-core. However, it should be noted that the results shown in Fig.\ \ref{fig:msrenc} depend heavily on the underlying assumptions, such as $\tau_{c,s} = \tau_c/2$. Furthermore, about 12\% fissions outside of the active core occurred in the MSRE \cite{Compere1975}; therefore, the ``effective'' core residence time is potentially different from the reported physical value. Finally, it is remarked that the choice of the degree of ex-core mixing significantly affects the relative DNP recirculation contribution to reactivity, ranging from $\sim$10\% for low mixing to more than $20\%$ for high mixing, see Fig.\ \ref{fig:msrenc}(right); nevertheless, recirculation plays a role at all levels of mixing. For comparison, the relative DNP recirculation contribution to reactivity predicted by the detailed system-code model from \cite{Fischer2023} is $\sim$18\%, indicating good agreement with the low in-core mixing/high ex-core mixing model presented here. However, the static reactivity loss was calculated as $\sim$37\% with the system-code model, thus pointing at the continuing existence of discrepancies, which will need to be reconciled in future work.

\begin{figure}
\centering
\includegraphics[width=0.7\linewidth]{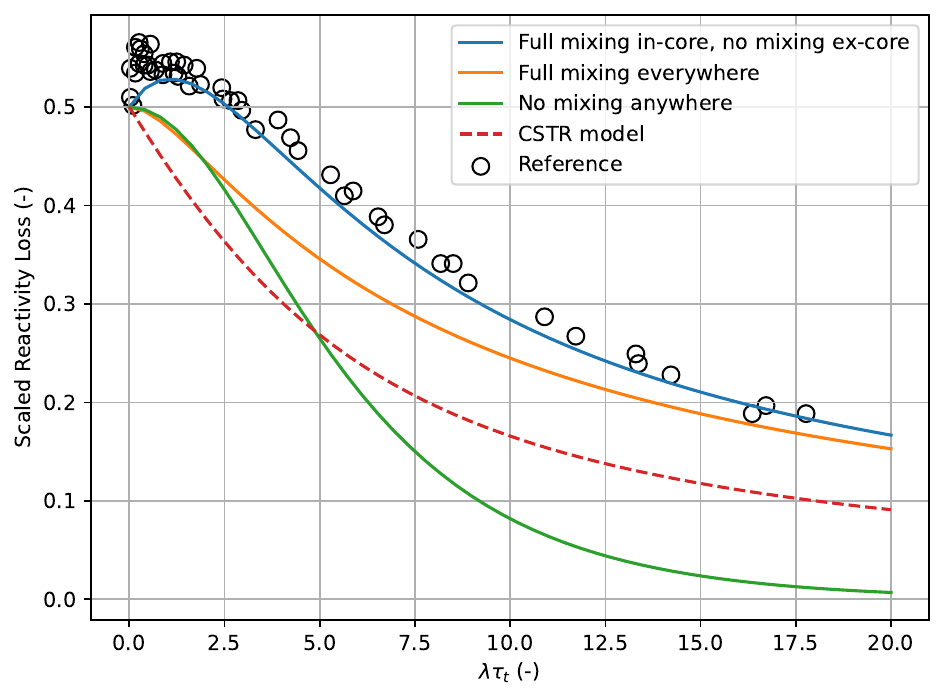}
\caption{Scaled reactivity loss (Eq.\ \ref{eq:scaled_loss}) as a function of $\lambda\tau_t$ for the EVOL reference MSFR for several cases of mixing in-core and ex-core compared to the conventional CSTR model (see Section \ref{sec:gamma}) and reference high-fidelity data obtained using Serpent 2 with precursor advection \cite{Aufiero2014}. In the figure, 1 = complete loss of DNPs. As a result of data extraction, error bars for the reference data are not plotted; the reader is referred to \cite{Aufiero2014} instead.}
\label{fig:evol_ind}
\end{figure}

\begin{figure}
\centering
\includegraphics[width=0.7\linewidth]{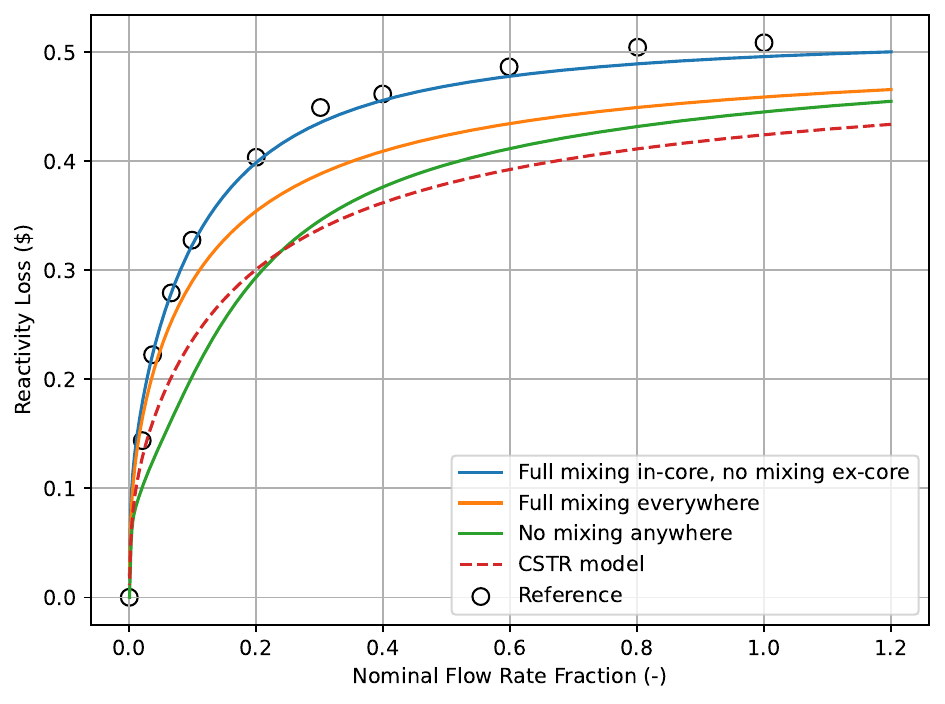}
\caption{Static reactivity loss (Eq.\ \ref{eq:k0_1}) as a function of flow rate relative to the nominal value for the EVOL reference MSFR for several cases of mixing in-core and ex-core compared to the conventional CSTR model (Eq.\ \ref{eq:cstr}) and reference high-fidelity data obtained using Serpent 2 with precursor advection \cite{Aufiero2014}. As a result of data extraction, error bars for the reference data are not plotted; the reader is referred to \cite{Aufiero2014} instead.}
\label{fig:evol_tot}
\end{figure}

In the second benchmark problem, the reference MSFR geometry from the EVOL project was considered. For this configuration, the static reactivity loss due to precursor drift, both for individual groups (i.e.\ as function of $\lambda\tau_t = \lambda(\tau_c+\tau_e)$) and overall as a function of flow rate were studied in \cite{Aufiero2014}. The reference EVOL geometry (referred to as ``k-epsilon case study'' in \cite{Aufiero2014}) features an axisymmetric cylindrical core with a non-uniform, turbulent velocity field with a large recirculation vortex as deduced from a k-epsilon RANS Computational Fluid Dynamics simulation. The ex-core velocity field, on the other hand, is rather uniform \cite{Aufiero2014}. In the presented framework, this would likely be indicative of substantial mixing in-core and limited mixing ex-core. Figure \ref{fig:evol_ind} shows the calculated scaled reactivity loss as a function of $\lambda\tau_t$ for several cases of mixing in-core and ex-core compared to the conventional CSTR model and reference high-fidelity data obtained using Serpent 2 with precursor advection \cite{Aufiero2014}. Very good agreement is obtained for the case when full mixing in-core and no mixing ex-core is considered: this is in accordance with expectation based on the geometry of the problem and the flow field used. Note that it appears that the increase of reactivity loss with $\lambda\tau_t$ for $\lambda\tau_t$ close to zero discussed above seems to be manifested also in the reference data, although their scatter as well as uncertainty (see \cite{Aufiero2014}) are substantial in the vicinity of $\lambda\tau_t=0$. The good agreement of the ``full mixing in-core and no mixing ex-core'' model is confirmed for the overall reactivity loss as a function of flow rate relative to the nominal value, see Fig.\ \ref{fig:evol_tot}. Eight DNP groups were used with the data taken from \cite{Aufiero2014} for $^{235}$U loading.

\begin{figure}
\centering
\includegraphics[width=0.7\linewidth]{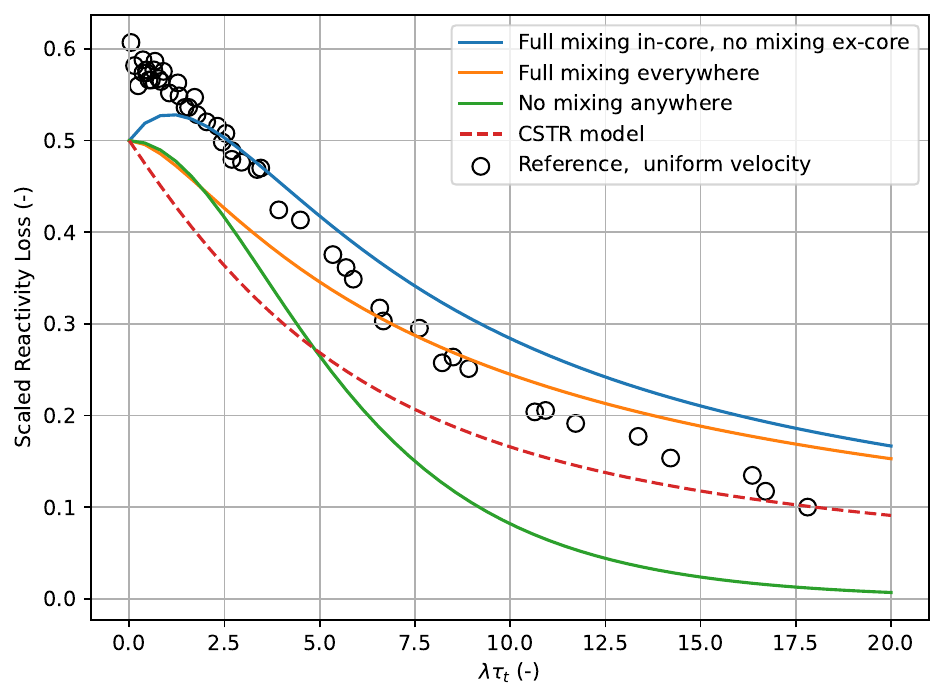}
\caption{Scaled reactivity loss (Eq.\ \ref{eq:scaled_loss}) as a function of $\lambda\tau_t$ for the EVOL MSFR with uniform velocity for several cases of mixing in-core and ex-core compared to the conventional CSTR model (see Section \ref{sec:gamma}) and reference high-fidelity data obtained using Serpent 2 with precursor advection \cite{Aufiero2014}. In the figure, 1 = complete loss of DNPs. As a result of data extraction, error bars for the reference data are not plotted; the reader is referred to \cite{Aufiero2014} instead.}
\label{fig:evol_ind_uni}
\end{figure}

In \cite{Aufiero2014}, extensive discussion is dedicated to the role of DNP spatial importance, which has been neglected in the current framework. Nevertheless, very good correspondence with the reference has been obtained. This could be explained as follows: to the extent the active core can be considered a perfectly-mixed volume, the creation and decay of flowing precursors essentially cannot be attributed to any spatial location. As a result, spatial importance effects are ``washed out''. According to the results shown in \cite{Aufiero2014}, perfect mixing is likely a good approximation, considering that even the distribution of the short-lived DNPs appears to be much smoother than the one-group flux \cite{Brovchenko2019}. This is corroborated by the fact that for the uniform-velocity case of \cite{Aufiero2014}, rather poor agreement was obtained (see Fig. \ref{fig:evol_ind_uni}); in such a situation, the spatial importance likely plays a more significant role. It is noted in passing that the ``full mixing in-core and no mixing ex-core'' model is in principle very similar to the CSTR model: the key difference being $\tau_{c,i} \neq \tau_{c,s}$.

Overall, the two benchmark problems indicate that the developed framework can be used to estimate the static reactivity loss of circulating-fuel reactors with reasonable accuracy.

\subsection{Zero-power transfer function}
\begin{figure}
\centering
\includegraphics[width=0.85\linewidth]{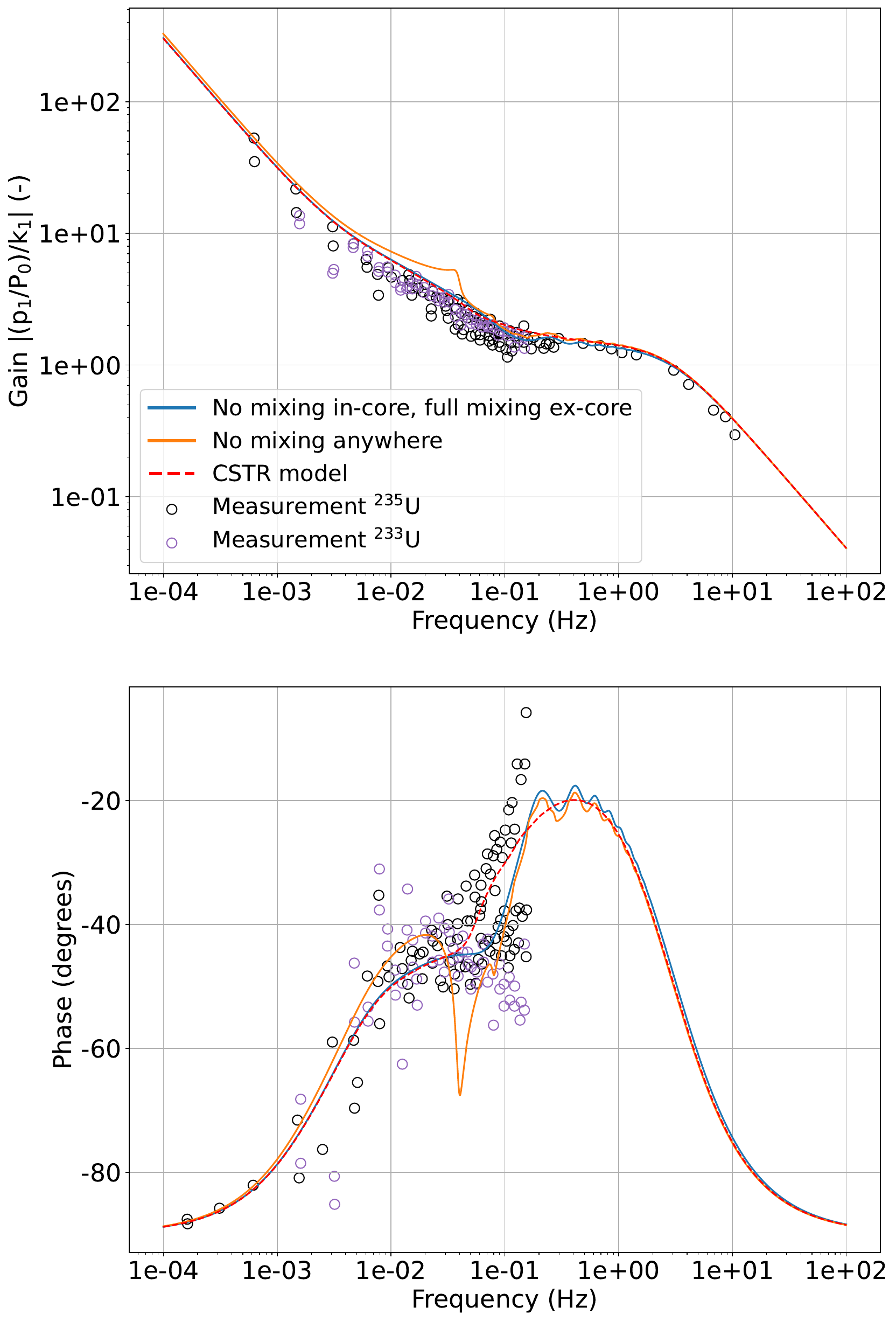}
\caption{Zero-power transfer function $Z(s)$ of the MSRE computed for several cases of mixing in-core and ex-core compared to the conventional CSTR model (Eq.\ \ref{eq:cstr_z}) and experimental data \cite{Prince1968, Steffy1970}. Top: amplitude, bottom: phase. Note that $s = 2\pi i f$ with $f$ being the frequency. The computed transfer functions assume kinetics data corresponding to
the $^{235}$U fuel loading of the MSRE (Table \ref{tab:msrednp}).}
\label{fig:msre_z}
\end{figure}

To the best of author's knowledge, there are no high-fidelity numerical results for zero-power transfer functions of flowing-fuel reactors. Therefore, the only available reference is the MSRE experimental data \cite{Prince1968, Steffy1970}. Using the same MSRE model inputs as in the previous section, the transfer functions were calculated for:
\begin{enumerate}
	\item The case of no mixing in-core and full mixing ex-core (i.e.\ the one with best agreement in the previous section), 
	\item The case with no mixing anywhere, i.e.\ a plug-flow approach similar to the one developed in \cite{Bures2024b},
	\item The CSTR model.
\end{enumerate}
It can be seen in Fig.\ \ref{fig:msre_z} that due to their scatter, the comparison with experimental data cannot distinguish between the ``no mixing in-core and full mixing ex-core'' model and the CSTR model. Nevertheless, the comparison at least indicates that the developed model is able to capture the dynamic response of the system well. Regarding the plug-flow approach, the clearly visible erroneous recirculation peak absent from reference experimental data suggests that mixing in the ex-core indeed needs to be captured in a successful model. This peak and its significance are further discussed in \cite{Bures2024b}.

\begin{figure}
\centering
\includegraphics[width=0.85\linewidth]{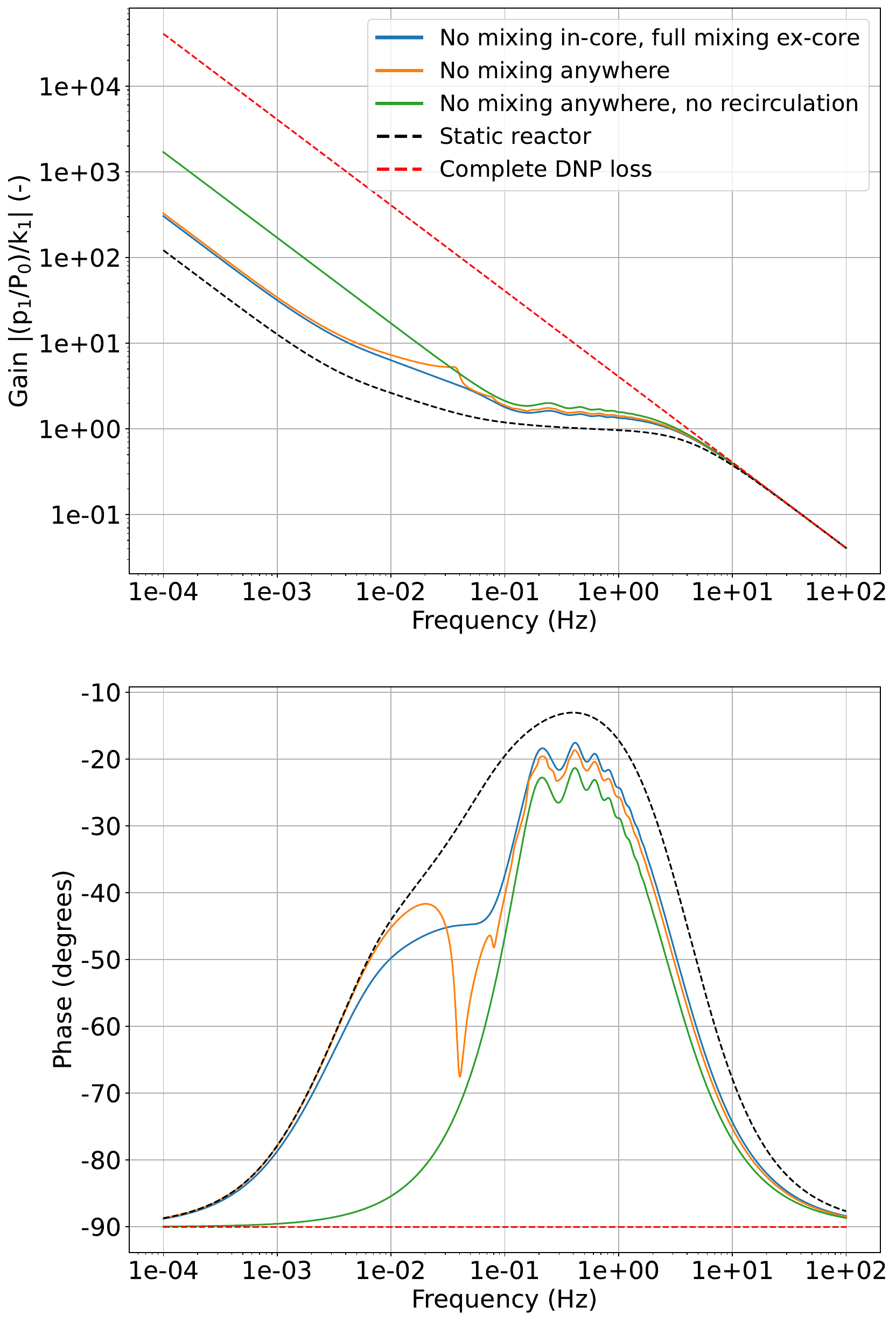}
\caption{Zero-power transfer function $Z(s)$ of the MSRE computed for several cases of mixing in-core and ex-core and recirculation. Zero-power transfer function of a static reactor without precursor drift and one with complete DNP loss are also shown for reference. Top: amplitude, bottom: phase. Note that $s = 2\pi i f$ with $f$ being the frequency. The computed transfer functions assume kinetics data corresponding to the $^{235}$U fuel loading of the MSRE (Table \ref{tab:msrednp}).}
\label{fig:msre_z_recirc}
\end{figure}

As a final study, the MSRE configuration was also evaluated with the plug-flow approach without recirculation with the results compared to the previously obtained solutions in Fig.\ \ref{fig:msre_z_recirc}. The overall increase in gain and decrease in phase for the circulating-fuel cases with respect to the transfer function of a static reactor can be observed; this is a direct effect of the precursor drift, making the reactor more responsive to perturbations. The lack of recirculation accentuates this behaviour even further, as DNPs are further removed from the active core. In the limit of complete DNP removal, $Z(s)$ would tend to $\beta/\Lambda/s$. Note that the lack of recirculation impacts the transfer function in yet another way: the prominent peak visible in the response with recirculation and its second harmonic disappear, confirming that they occur precisely due to presence of recirculation.

\section{Conclusion}
\label{sec:conc}
This study developed and exercised a residence-time-theory framework to quantify how delayed-neutron-precursor (DNP) drift and recirculation shape the static reactivity loss and zero-power dynamic response of circulating-fuel reactors (CFRs). By representing in-core and ex-core residence-time distributions with gamma distributions, the resulting theoretical model captured the full spectrum from perfect mixing to plug-flow behaviour within a closed-form set of expressions for the reactivity loss, the recirculation share, and the zero-power transfer function. The derived formulas were shown to reduce to the classical Continuous-Stirred-Tank-Reactor (CSTR) and plug-flow limits, while being able to capture intermediate mixing regimes also. To the author's knowledge, this is the first time such a modelling approach to DNP dynamics has been used, in particular when the generalisation to gamma distributions is considered.

The developed model was first used to assess the impact of DNP decay constant and system residence times on reactivity loss and recirculation fraction. It was found that, in general, the recirculation has the highest impact when the ex-core and in-core residence times are comparable and the product of the DNP decay constant $\lambda$ and the in-core residence time $\tau_c$ is small. Subsequently, an MSRE-like system was examined in more detail and a surprising phenomenon leading to the increase of reactivity loss with a decrease of flow rate was found to occur under certain conditions. It was credited to the competing effects of in-core and ex-core DNP decay on the reactivity loss. This phenomenon was also found to be fully dependent on the occurrence on recirculation of DNPs in the system. 

The reactivity-loss analysis was concluded by two benchmark studies. First, it was found that for the MSRE-like system, in order to match the experimentally-reported static reactivity loss, high degree of mixing should be assumed ex-core and low in-core, respectively. This was considered to be in-line with the physical design of the MSRE but in conflict with the conventional CSTR assumption of the opposite arrangement. With this modelling choice, recirculation was found to be significant with about 20\% of DNP reactivity being due to recirculation back to the core under steady-state conditions. This result agreed well with detailed system-code results, even though a discrepancy in static reactivity loss was found. For the second benchmark problem, the EVOL reference Molten-Salt Fast Reactor, high-fidelity Serpent-2 calculations were closely matched when full mixing was assumed in-core and plug flow ex-core, mirroring the effects of a highly turbulent active core and comparatively uniform ex-core flow observed in Computational Fluid Dynamic modelling. Conversely, for a uniform flow in-core, poor agreement was found, likely due to increased role of spatial importance in that scenario.

The static reactivity loss investigation was complemented by reproducing the frequency-domain zero-power transfer function of the MSRE with good agreement, again assuming high degree of mixing ex-core and low in-core, although the scatter of experimental data prevented meaningful differentation against the conventional CSTR model. The no-mixing (plug-flow) modelling lead to the occurrence of an erroneous recirculation peak in the frequency response, indicating the importance of correctly capturing fluid mixing in the model.

In future work, the presented framework will be applied to other CFR designs, either to available experimental data or numerical calculations. In particular the Molten Salt Research Reactor \cite{Abilene2022} or the Molten Chloride Reactor Experiment \cite{Walls2023} could be promising reference designs, since their construction will likely motivate high-fidelity simulations and lead to future experimental measurements. The recently-developed thermal molten-salt reactor benchmark \cite{Pfahl2025} will also be a valuable test case, since it features no mixing either in-core or ex-core, a situation not studied in the present paper. Further model development will focus on the inclusion of thermal effects and the derivation of transfer functions under full power conditions. Another avenue for future investigation are for example the introduction of more than two control volumes into the framework and characterisation of residence-time distributions of various reactors, this being a necessary input to the presented models. The ideas sketched out in the \hyperref[sec:app]{Appendix} will also be developed further.

\section*{Selected nomenclature}
\subsection*{Symbols}
\noindent
$C$: precursor concentration\newline
$E$: residence-time or life-expectancy distribution\newline
$k$, $\rho$: reactivity in dollars and in pcm\newline
$P$, $p$: fission rate, its perturbation\newline
$S$: precursor source\newline
$Q$: volumetric flow rate\newline
$Z$: zero-power transfer function\newline
$\beta$: delayed neutron fraction; no subscript indicates total\newline
$\Lambda$: prompt neutron generation time\newline
$\lambda$: precursor decay constant\newline
$\tau$: age\newline
$\Omega$, $\omega$: integral of concentration of precursors over the active core, its perturbation
\subsection*{Subscripts}
\noindent
$0$: steady-state\newline
$c$: core\newline
$e$: ex-core\newline
$\textit{i}$: inflow\newline
$\textit{in}$: inlet\newline
$\textit{out}$: outlet\newline
$s$: source\newline
$t$: total

\section*{Declaration of competing interest}
The author declares a competing interest due to affiliation with Saltfoss Energy ApS, a company actively involved in the development of circulating-fuel, molten-salt reactor technology, in the form of employment.

\section*{Acknowledgements}
The author would like to thank Lorenz Fischer and Jacob Groth-Jensen for valuable suggestions regarding the submitted manuscript.

\clearpage
\appendix

\phantomsection
\section*{Appendix\quad Possible extensions of the age-based framework}
\label{sec:app}

As presented in this paper, two deficiencies of the age-based framework can be identified:
\begin{enumerate}
    \item Neglecting of DNP importance in the calculation of $\Omega(t)$ in Eq.\ \ref{eq:power_pk}.
    \item Lack of direct applicability of the framework in the time domain.
\end{enumerate}
Conversely, a key insight was obtained in the sense that this framework reduces the dimensionality of advection from three spatial
dimensions to one age dimension. This insight could potentially be leveraged to resolve both of the issues listed above. In principle, knowledge of flow patterns in the control volume $\mathcal{V}$, such as the reactor core, would allow for constructing a map between the 3D spatial coordinate vector \( \mathbf{x} \) to the scalar age $\tau$, i.e.\ \( \tau = f(\mathbf{x}) \) for some function \( f \). Such a map \( \mathbf{x} \to \tau \) is in general not a bijection, meaning that each value of \( \tau \) would correspond to a two-dimensional level set \( S_{\tau} \) in \( \mathbb{R}^3 \).

Then, considering a shape function (e.g.\ DNP importance) satisfying the normalization condition:
\begin{equation}
	\int_\mathcal{V} g(\mathbf{x}) \, d^3\mathbf{x} = 1,
\end{equation}
the use of the \textit{coarea formula} \cite{Federer1969} would allow the 3D integral to be expressed as an integral over the scalar \( \tau \) and the corresponding level sets \( S_{\tau} \). The coarea formula in this context reads:
\begin{equation}
	\int_\mathcal{V} g(\mathbf{x}) \, d^3\mathbf{x} = \int_0^\infty  \left( \int_{S_{\tau} \cap \mathcal{V}} \frac{g(\mathbf{x})}{|\nabla f(\mathbf{x})|} \, dS \right) d\tau,
\end{equation}
where:
	\begin{itemize}
		\item \( \nabla f(\mathbf{x}) \) is the gradient of the map \( f(\mathbf{x}) \), and
		\item \( dS \) is the surface measure on the level set \( S_{\tau} \).
	\end{itemize}
The normalization condition translates into ensuring that the 1D integral over \( \tau \) satisfies:
\begin{equation}
	\int_0^\infty h(\tau) d\tau = 1,
\end{equation}
where the function \( h(\tau) \) is given by:
\begin{equation}
	h(\tau) = \int_{S_{\tau} \cap \mathcal{V}} \frac{g(\mathbf{x})}{|\nabla f(\mathbf{x})|} \, dS.
\end{equation}
If one were able to use the map $\tau = f(\mathbf{x})$ to derive an importance-weighted persistence kernel $A_g(t,t-\tau)$ (cf.\ Eq.\ \ref{eq:omega}), the framework could be extended beyond cases with uniform DNP importance.

Coming now to the second point, i.e.\ to application of the framework in the time domain, here the key idea could be that rather than using an integral, black-box inlet/outlet representation of control volumes (e.g.\ Eq.\ \ref{eq:outlet_core}), one could use a differential advection-like framework. In this framework, one can write a governing for DNP concentration per unit age $\xi$ in the active core $c$ as:
\begin{equation}
    \frac{\partial \xi(t,\tau)}{\partial t} + \frac{\partial \xi(t,\tau)}{\partial \tau} = q(t,\tau) -r(\tau)\xi(t,\tau) - \lambda \xi(t,\tau).
    \label{eq:hazard_eq}
\end{equation}
Here $\xi(t,\tau)$ is the DNP concentration loading in fluid parcels of age $\tau$ at time $t$. Additionally, $r(\tau)$ is the fluid exit-hazard \cite{Rodrigues2021, Cox1984}:
\begin{equation}
    r(\tau) = \frac{E_c(\tau)}{1-\int_0^\tau E_c(t')dt'},
\end{equation}
and $q(t,\tau)$ is the injection of DNPs into the age class $\tau$, which could be obtained from the power shape function and the aforementioned map $\tau = f(\mathbf{x})$. The boundary condition at age $\tau=0$ is:
\begin{equation}
    \xi(t,0) = \frac{Q}{V_c}C_{\textit{in}}(t) = \frac{1}{\tau_c}C_{\textit{in}}(t).
\end{equation}
This formulation could be used as a foundation for solving the dynamics of the DNP population, e.g.\ after the governing equation \eqref{eq:hazard_eq} is discretized using a finite number of age classes.

For an illustration, one can consider a one-dimensional plug-flow system with velocity $U$ and length $L_c$, for which the governing equation for DNPs in group $j$ in the active core is given as \cite{Bures2024b}:
\begin{align}
    \frac{\partial C_j(t,z)}{\partial t} + U\frac{\partial C_j(t,z)}{\partial z} &= \frac{\beta_j}{\Lambda}\chi(z)P(t) - \lambda C_j(t,z), \ \ 0 < z < L_c, \label{eq:pfr_eq} \\
    C_j(t,0) &= C_{\textit{in}}(t).
\end{align}
For this system, the map $f(\mathbf{x})$ can be obtained in a straightforward manner as $\tau=z/U$ and the relation between $C_j$ and $\xi_j$ is:
\begin{equation}
    \xi_j(t,\tau) = \frac{1}{\tau_c}C_j(t,U\tau).
\end{equation}
Evidently, Eq.\ \ref{eq:pfr_eq} is equivalent to Eq.\ \ref{eq:hazard_eq} with the injection rate of $\beta/\Lambda\cdot\chi(U\tau)P(t)$ and the hazard rate given in the distributional sense as $\delta(\tau-\tau_c)$, i.e.\ all DNPs leaving the core instantaneously at age $\tau_c=L_c/U$ consistently with a Dirac residence-time distribution \cite{Rodrigues2021}.

For a second illustration, one can consider the other extreme case, i.e.\ a perfectly mixed system with an exponential residence-time distribution. Then, the hazard rate is constant and equal to $\tau_c$; subsequently, Eq.\ \ref{eq:hazard_eq} can be integrated for the DNP group $j$ over all ages and multiplied by volume $V_c$ to obtain:
\begin{equation}
      \frac{d \Omega_j(t)}{d t} + V_c \big[\xi_j(t,\infty) - \xi_j(t,0)\big] = S_j(t) -\frac{1}{\tau_c}\Omega_j(t)- \lambda_j \Omega_j(t).  
\end{equation}
Since $\xi_j(t,\infty)=0$ and $V_c \xi_j(t,0) = QC_{\textit{in}}(t)$:
\begin{equation}
      \frac{d \Omega_j(t)}{d t} = S_j(t) -\frac{1}{\tau_c}\Omega_j(t)- \lambda_j \Omega_j(t) + QC_{\textit{in}}(t),
\end{equation}
which is the global material balance for precursors in a CSTR active core control volume \cite{Cammi2010,Bures2024b} with $S_j$ given by Eq.\ \ref{eq:sj}.

\clearpage
\printbibliography

\end{document}